\documentclass[epj,twocolumn]{webofc}
\usepackage[varg]{txfonts}   
\usepackage{graphicx}

\newcommand{\nat}{Nature,~}
\newcommand{\mnras}{MNRAS,~}

\newcommand{\apj}{ApJ,~}
\newcommand{\apjl}{ApJL,~}
\newcommand{\apjs}{ApJ Suppl. Series,~}
\newcommand{\aap}{A\&A,~}

\newcommand{\pasj}{PASJ,~}

\wocname{epj}
\woctitle{Seismology of the Sun and the Distant Stars 2016}
\begin{document}
\title{Insights from Asteroseismology of Massive Stars:
The Need for Additional Angular Momentum Transport Mechanisms}
%

\author{
\firstname{Ehsan} \lastname{Moravveji}\inst{1}\fnsep\thanks{\email{Ehsan.Moravveji@kuleuven.be}} 
}

\institute{Institute of Astronomy, KU Leuven, Celestijnenlaan 200D, B-3001 Leuven, Belgium
          }

\abstract{
In massive stars, rotation and oscillatory waves can have a tight interplay.
In order to assess the importance of additional angular momentum transport 
mechanisms other than rotation, we compare the asteroseismic properties of a 
uniformly rotating model and a differentially rotating one.
Accordingly, we employ the observed period spacing of 36 dipole g-modes in the Kepler 
$\sim3.2$\,M$_\odot$ target KIC\,7760680 to discriminate between these two models.
We favor the uniformly rotating model, which fully satisfies all observational constraints.
Therefore, efficient angular momentum transport by additional mechanisms such as 
internal gravity waves, heat-driven modes and magnetic field is needed during
early main sequence evolution of massive stars.
}
\maketitle
%
\section{Introduction and objective}\label{s-intro}

Massive stars are moderate to fast rotators \cite{huang-2010-01}. 
Rotation modifies the thermal and structural equilibrium in stars,  
launches meridional circulation currents \cite{zahn-1992-01}, and 
triggers a handful of hydrodynamical instabilities (Maeder 2009, 
\cite{maeder-2009-book}).
Stars with masses exceeding $\sim1.3$\,M$_\odot$ have diminishing 
outer convective envelopes, and
above $\sim2$\,M$_\odot$ the envelope becomes predominantly radiative 
(except for the narrow helium and/or iron opacity bumps).
The radiative envelope provides a suitable propagation cavity for pressure 
(p-) and gravity (g-) modes to traverse it back and forth roughly on the 
dynamical timescale $\tau_{\rm dyn}$.
Because the nuclear timescale $\tau_{\rm nuc}$ is much longer than the dynamical
timescale, the oscillatory motions in stars have sufficient time to interact 
with the background star.
One critical interaction is an efficient redistribution of angular 
momentum inside stars by non-radial oscillations 
\cite{ando-1983-01,unno-1989-book,lee-1993-01,townsend-2009-01,lee-2016-01}, 
or through internal gravity waves \cite{rogers-2013-01,rogers-2015-01}.
Townsend (2009, \cite{townsend-2009-01}) has demonstrated that in a 5M$_\odot$ 
star, a very efficient angular momentum transport in the envelope takes place 
in less than 1\,000 years.
The coupling between the stellar structure and the pulsations is non-linear, 
and forms an interdependency:
the oscillations can modify the structure on short timescales and the structure
determines the oscillation properties.
Therefore in massive stars, the coexistence of rotation and pulsation can 
have profound consequences and can impact the global observables
(e.g. abundances, surface rotation, mass loss rate).

The conservation of angular momentum determines the coupling between 
the vertical component of the angular velocity $\Omega(r)$, the velocity of the
advective meridional circulation currents $U$, and the additional angular momentum flux
carried by waves $\mathcal{F}$ 
(see e.g. \cite{chaboyer-1992-01,zahn-1992-01,maeder-2009-book} for more 
details)

\begin{align}\label{e-AM}
\frac{\partial}{\partial t}\left(\rho r^2\Omega\right) &= 
\frac{1}{5r^2}\frac{\partial}{\partial r}\left(\rho r^4\Omega U\right)+
\frac{1}{r^2}\frac{\partial}{\partial r}\left(\rho\,\nu\,r^4\frac{\partial\Omega}{\partial r}\right) \nonumber \\
&-\frac{3}{2r^2}\frac{\partial}{\partial r}(r^2\,\mathcal{F}), 
\end{align}
where, $\nu$ is the vertical turbulent diffusivity.
In some stellar structure and evolution models (e.g. Heger et al. 2000, 
\cite{heger-2000-01}), the advective term is approximated as a diffusion process.
Through Eq.\,(\ref{e-AM}), the angular momentum flux $\mathcal{F}$ can be extracted 
from one part of the star, and deposited in a different region.
Consequently, some layers inside stars spin up, while some other regions spin down, 
trying to conserve the net amount of angular momentum.
Eventually, some angular momentum will be lost through wind mass loss and/or via 
binary interaction.

How significantly can the waves contribute to the angular momentum transport in 
rotating stars?
To answer this quantitatively, one must compare models that incorporate 
the last term in Eq.\,(\ref{e-AM}), and exploit the observable consequences 
\cite{ando-1983-01,lee-2016-01}.
In this study, however, we present a preliminary and {\it qualitative} comparison
between two models that represent two possible extremes, without any direct 
computation of extra angular momentum flux $\mathcal{F}$.
One model is forced to rotate uniformly (abbreviated as UR), and it
represents a case where waves contribute very efficiently to maintain
rigid rotation.
The other model has a differential rotation (abbreviated as DR).
The latter represents ignoring the angular momentum flux carried by waves 
$\mathcal{F}=0$, and obeys the standard angular momentum transport by rotation 
in a diffusion approximation \cite{pinsonneault-1989-01,heger-2000-01}.

The high-precision Kepler observations of pulsating stars has opened a unique
window to empirically constrain the angular momentum distribution inside
heat-driven pulsators.
This facilitates putting our knowledge of angular momentum transport, and the 
underlying assumptions under scrutiny.
As a demonstration, the radiative envelopes of three evolved $\gamma$\,Dor stars 
-- near the end of core H burning phase -- were shown to rotate almost uniformly 
\cite{kurtz-2014-01,saio-2015-01,murphy-2016-01}.
The descendant of these stars are the red giants which are also shown to have a 
core-to-surface rotation rate slower than the model predictions \cite{beck-2012-01}.

In this paper, we concentrate on the moderately rotating and slowly pulsating 
B-type star KIC\,7760680.
Our stellar models are computed using MESA \cite{paxton-2011-01}, which incorporate 
the effect of rotation \cite{paxton-2013-01}. 
The theoretical adiabatic frequencies are computed using GYRE \cite{townsend-2013-01}, 
which treats the effects of rotation on pulsations through the Traditional Approximation 
of Rotation (TAR) \cite{bouabid-2013-01}.
Our inferences regarding model UR is heavily based on the in-depth seismic modeling presented 
by Moravveji et al. (2016, hereafter Paper1, \cite{moravveji-2016-02}).
We explicitly try to address two questions, in the light of strict observational 
(asteroseismic and spectroscopic) constraints that we place on this star.
(Q1) Which of the rigidly rotating (UR) or differentially rotating (DR) models fulfil 
     the observations?
(Q2) What is the minimum required vertical diffusive mixing to maintain (near) rigid 
     rotation?

\section{Observational constraints}\label{s-obs}
KIC\,7760680 was thoroughly studied by P{\'a}pics et al. (2015, \cite{papics-2015-01}).
From spectroscopy the projected surface rotation velocity is 
$v\sin i=62\pm5$ (km sec$^{-1}$).
This is the first requirement that we try to fulfil in model UR and DR.
In Paper1, stringent constraints were placed on the global and mixing 
parameters of this star through forward seismic modeling.
In short, KIC\,7760680 is a $\sim3.25$\,M$_\odot$ B8V star, with core hydrogen 
mass fraction $X_{\rm c}=0.503\pm0.001$.
{\rm Assuming a solid body rotation}, and trying to fulfil the observational 
constraints during asterseismic modeling, the surface rotation frequency 
is tightly bound to 
$\Omega_{\rm surf}=2\pi f_{\rm rot}=3.0096^{+0.0415}_{-0.0591}$ radian day$^{-1}$.

KIC\,7760680 exhibits $\mathcal{N}_{\rm obs}=36$ consecutive dipole prograde g-modes.
The first and the last mode in this series have periods of 
$P_1=0.86930$ and $P_{36}=1.46046$ days, respectively.
The $\mathcal{N}_{\rm obs}=36$ is the second asteroseismic constraint that we 
strictly require in this study.
The motivation for this requirement is the high sensitivity of inertial g-modes
to the rotation rate (see e.g. Fig.\,2 in Townsend 2005, \cite{townsend-2005-01}).

\section{Uniform versus differential rotation}\label{s-UR-DR}

\begin{figure}[t]
\includegraphics[width=\columnwidth]{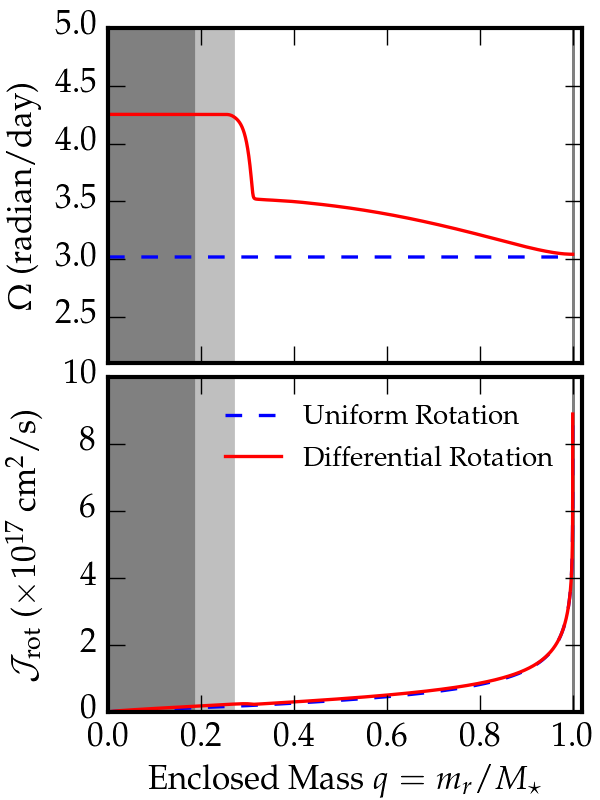}
\caption{Comparison between the uniformly (UR) and differentially (DR) rotating models.
The former is shown with a dashed blue line, and the latter with a solid red curve.
Table\,\ref{t-models} lists their global parameters. 
(Top) The internal angular velocity versus enclosed mass. 
Both models have identical radius and surface rotation velocity.
(Bottom) The internal angular momentum distribution $\mathcal{J}_{\rm rot}$ versus
enclosed mass.
The model DR has a slightly higher angular momentum across the interior.
}
\label{f-J}
\end{figure}

To attack the first question Q1, we construct two MESA models (UR and DR) 
that have very similar stellar mass, radius, mass of the convective core 
$M_{\rm cc}$ and center hydrogen mass $X_{\rm c}$.
Table\,\ref{t-models} summarises the parameters of both models.
Based on Paper1, the model UR was computed from non-rotating MESA tracks, and the uniform 
rotation was only imposed during the GYRE computations of adiabatic frequencies.
Model DR starts off the main sequence with an initial surface rotation velocity 
$v_{\rm ini}=75$\,km\,sec$^{-1}$.
The evolution of model DR is terminated at $X_{\rm c}=0.495$, and the surface rotation
velocity has evolved to 68\,km\,sec$^{-1}$.
Thus, both UR and DR satisfy the first requirement in Section\,\ref{s-obs}.

\begin{table}
\centering
\caption{The physical parameters of the models UR and DR.
OV, Diff and Rot stand for core overshoot mixing, vertical diffusive
mixing and rotational mixing, respectively.
$v_{\rm rot}$ is the initial rotation rate at zero-age-main-sequence.}
\label{t-models}
\begin{tabular}{lll}
\hline
  & Model UR & Model DR \\
\hline \hline
MESA      & non-rotating & rotating \\
ZAMS $v_{\rm rot}$ (km\,sec$^{-1}$) & 0 & 75 \\
GYRE      & TAR & TAR \\
Mixing    & OV $+$ Diff. & OV $+$ Rot. \\
Star mass (M$_\odot$) & 3.25 & 3.20 \\
Core mass (M$_\odot$) & 0.621 & 0.606 \\
X$_{\rm c}$ & 0.503 & 0.495 \\
Radius (R$_\odot$) & 2.79 & 2.80 \\
\hline 
\end{tabular}
\end{table}

Fig.\ref{f-J} compares model UR (dashed line) and model DR (solid line).
The extent of the convective core (dark gray), and the core overshooting 
region (light gray) are highlighted.
The top panel shows the vertical angular velocity $\Omega(r)$. 
Model UR is assumed (and enforced) to rotate uniformly, whereas model DR 
rotates differentially.
Both models have roughly equal surface angular velocity, and radius 
(Table\,\ref{t-models}), hence similar surface rotation velocity.
Due to the efficient core boundary mixing by overshooting, this narrow
region strongly contributes to the transport of angular momentum from 
the core region to the envelope.
Thus, the narrow overshooting layer rotates synchronously with the core.
The bottom panel shows the specific angular momentum 
$\mathcal{J}_{\rm rot}=i_{\rm rot}\Omega$, where 
$i_{\rm rot}\approx2r^2/3$ is the specific moment of inertia.
Given the faster inner rotation of the model DR, it has slightly higher 
angular momentum $\mathcal{J}_{\rm rot}$.
This important feature is discussed in Section\,\ref{s-conc}.

\section{The asteroseismic test}\label{s-dP}

\begin{figure}[t]
\centering
\includegraphics[width=\columnwidth]{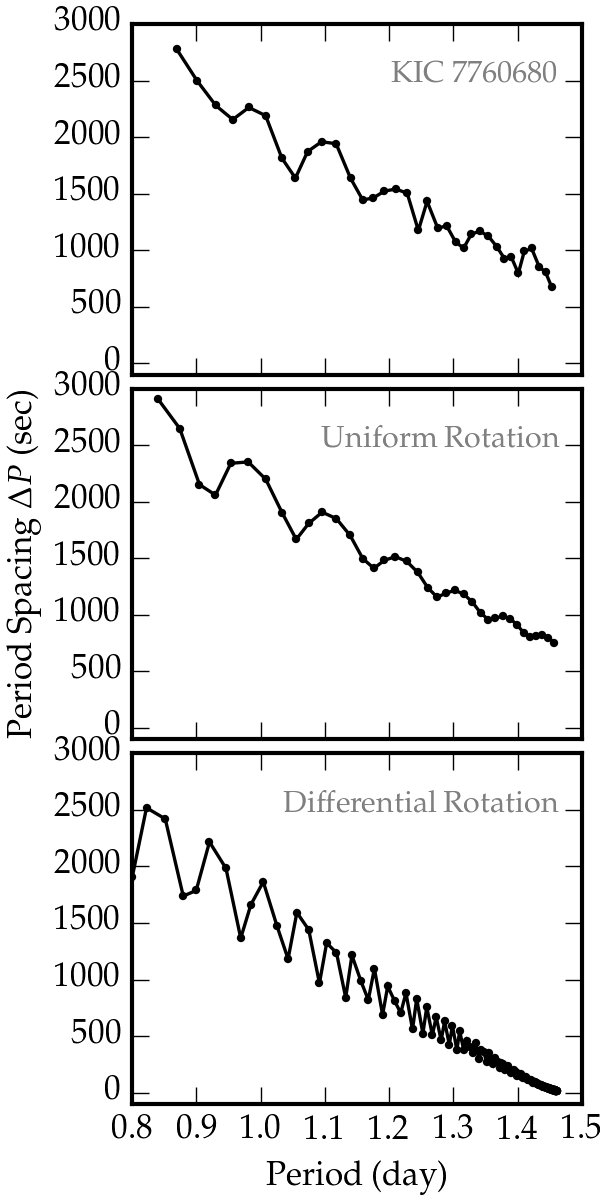}
\caption{The period spacing versus mode period.
(Top) The observed period spacing of KIC\,7760680 forming a series of 
$\mathcal{N}_{\rm obs}=36$ consecutive g-modes with period in the range 0.86930 and 1.46046 days.
(Middle) The period spacing from model UR with $\mathcal{N}_{\rm UR}=36$.
The properties of this model are elaborated in \cite{moravveji-2016-02}.
This model satisfies both observational constraints (Section\,\ref{s-obs}).
(Bottom) The period spacing from model DR. 
This model has $\mathcal{N}_{\rm DR}=156$ modes in the observed period range, which is 
incompatible with observations.
\label{f-dP}}
\end{figure}

To answer the first question Q1, we compute the period spacing for both models 
UR and DR, and verify if they fulfil the second requirement in Section\,\ref{s-obs}, 
i.e. $\mathcal{N}_{\rm obs}=36$ between $P_1$ and $P_{36}$.
Fig.\,\ref{f-dP} compares the period spacing $\Delta P_n=P_{n+1}-P_n$ from 
observation (top) with that of model UR (middle) and that of model DR (bottom).
Model UR is the best seismic model of KIC\,7760680 (Paper1), and it reproduces the 
number of observed modes $\mathcal{N}_{\rm UR}=\mathcal{N}_{\rm obs}$, in addition 
to the slope and the morphology of the observed period spacing.
Thus, model UR satisfies both observational requirements in Section\,\ref{s-obs}.

Model DR, however, has $\mathcal{N}_{\rm DR}=156$ modes between $P_1$ and $P_{36}$, which 
strongly contradicts with the observations (Fig.\,\ref{f-dP} top).
The period spacing of the very high-order g-modes in this model approach zero, whereas
the observed $\Delta P$ for the last mode in the series is 678\,sec.
Furthermore, the morphology of the period spacing pattern in this model does not 
resemble that of KIC\,7760680, due to strong differences with the observed one.
The fact that model DR does not fulfil the second observational requirement disqualifies
it from representing KIC\,7760680.

The current model of the diffusion of angular momentum employs
various ad-hoc efficiency factors \cite{pinsonneault-1989-01,heger-2000-01}.
These factors are poorly calibrated against observations, and are designed 
to suppress or boost the role of non-rotating and rotating mixing 
coefficients in transporting the angular momentum through the diffusion coefficient 
$\nu$ in Eq.\,(\ref{e-AM}).
For choosing model DR, we varied most of the efficiency factors up to 5 
orders-of-magnitude.
However, the general conclusion did not significantly change; 
thus, we set the efficiency factors to the default value of unity.
Furthermore, we also selected models closer to the zero-age-main sequence 
$X_{\rm c}\gtrsim0.60$ with slower initial rotation velocity $v_{\rm rot}\geq60$
km\,sec$^{-1}$.
For such models, $\mathcal{N}_{\rm DR}$ approaches $\mathcal{N}_{\rm obs}$.
But on the flip side, the morphology of the period spacing (the number and position of
dips) never resembled the observations.
This is expected.
The less-evolved models cannot develop extended $\mu$-gradient layers on top of 
their receding convective cores;
thus, the period spacings of models close to the zero-age-main-sequence hardly
deviate from the asymptotic period spacing (straight line).
In conclusion, we have so far not found any differentially rotating model
that respects the observational constraints in Section\,\ref{s-obs}, in addition to
providing a reasonable match to the observed period spacing.
As a result, we favor model UR, or any other model with near-uniform rotation as
the more likely representative of the internal distribution of angular
momentum in KIC\,7760680. 

\section{Discussion and conclusions}\label{s-conc}

Let's revisit Fig.\,\ref{f-J}.
The difference between the angular momentum profile
$\mathcal{J}_{\rm rot}$ in both models is minute, whereas their vertical 
angular velocities $\Omega(r)$ are different. 
In order to reduce the gradient of angular velocity in model DR, a small
amount of additional angular momentum must be transferred from the near-core region 
to the envelope.
Furthermore, small amount of angular momentum must be deposited from the star,
to spin down the core to the same level of the envelope, and allow the total angular
momentum $\mathcal{J}_{\rm rot}$ to approach that of model UR.

The need for additional angular momentum flux implies that $\mathcal{F}\ne0$ in 
Eq.\,(\ref{e-AM}).
Among possible mechanisms, other than rotation, that can actively redistribute 
angular momentum within
short timescales (compared to nuclear timescale) are the transport by internal
gravity waves \cite{talon-2005-01,rogers-2013-01,rogers-2015-01} and 
heat-driven nonradial modes \cite{ando-1983-01,lee-1993-01,townsend-2009-01}.
In fact, such additional angular momentum transport mechanisms are needed to
explain the near-uniform rotation in the three slowly rotating 
Kepler $\gamma$\,Dor pulsators KIC\,11145123 \cite{kurtz-2014-01}, KIC\,9244992
\cite{saio-2015-01} and KIC\,7661054 \cite{murphy-2016-01}, in addition
to the slowly rotating and pulsating B star KIC\,10526294 
\cite{papics-2014-01,moravveji-2015-01,triana-2015-01}.

We now turn to the address how much turbulent diffusivity is needed in the 
radiative envelope to sustain uniform rotation (Q2 in Section\,\ref{s-intro})?
In Paper1, we constrained the extra diffusive mixing above the overshooting zone
to $\log D_{\rm ext}=0.75\pm0.25$ cm$^2$\,sec$^{-1}$.
This value is between 100 to 1\,000 times less than the rotationally-induced 
mixing coefficients due to e.g. shear, meridional circulation and other proposed 
mechanisms in \cite{heger-2000-01} and \cite{maeder-2013-01}.
Therefore, in massive stars, small amount of turbulent diffusivity is sufficient
to maintain near-uniform rotation in their radiative envelopes.
This is understandable in the light of Eq.\,(\ref{e-AM});
positive additional angular momentum flux $\mathcal{F}>0$ would reduce turbulent
diffusivity $\nu$ needed to sustain rigid rotation.

Asteroseismology of massive stars offers a great opportunity to exploit the inner
workings of massive stars, which are inaccessible to us through other means.
This encourages to direct the future research to provide quantitative answer to
questions such as (a) how do oscillatory waves engage in angular momentum redistribution?
and (b) how much turbulent diffusivity is needed to explain pulsation behaviour
of stars, and maintain near-uniform rotation?

\vspace{2mm}
\noindent{\bf Acknowledgement.}
{\small 
E.~Moravveji is grateful to the organizers of the KASC9/TASC2 conference.
The research leading to these results has received funding from the People Programme (Marie
Curie Actions) of the European Union's Seventh Framework Programme FP7/2007-2013/ under REA
grant agreement N$^\circ$\,623303 (ASAMBA).}

%
%

\end{document}